\documentclass[aps,pra,reprint,superscriptaddress,longbibliography]{revtex4-1}

\usepackage{amsmath,amssymb}
\usepackage{mathrsfs}
\usepackage{graphicx}
\usepackage{upgreek}
\usepackage{color}
\usepackage[english]{babel}

\newcommand{\cL}{{\mathcal L}}

\newcommand{\cR}{{\mathcal R}}

\newcommand{\sA}{{\mathsf A}}
\newcommand{\sB}{{\mathsf B}}
\newcommand{\sH}{{\mathsf H}}
\newcommand{\sJ}{{\mathsf J}}
\newcommand{\sL}{{\mathsf L}}
\newcommand{\sP}{{\mathsf P}}
\newcommand{\sR}{{\mathsf R}}
\newcommand{\sS}{{\mathsf S}}
\newcommand{\sT}{{\mathsf T}}

\newcommand{\pop}{{\mathsf p}}
\newcommand{\pOmega}{{\mathsf{\Omega}}}
\newcommand{\eps}{\varepsilon}

\newcommand{\matel}[3]{{\left\langle \vphantom{#1 #2 #3} #1 \,\right\vert
\left.
 \hspace{-0.15em} \vphantom{#1 #2 #3} #2 \,\right\vert \left.
 \hspace{-0.15em} \vphantom{#1 #2 #3} #3\right\rangle}}
\newcommand{\ketbra}[2]{\vert #1 \rangle \langle #2 \vert}
\newcommand{\be}{\begin{equation}}
\newcommand{\ee}{\end{equation}}
\newcommand{\bea}{\begin{eqnarray}}
\newcommand{\eea}{\end{eqnarray}}
\newcommand{\dint}[1]{\mathrm{d} #1 ~}
\newcommand{\mitl}[1]{\left \langle #1 \right \rangle}

\begin{document}
 
\title{Spatio-Orientational Decoherence of Nanoparticles}

\author{Benjamin A. Stickler}
\affiliation{University of Duisburg-Essen, Faculty of Physics, Lotharstra\ss e 1, 47048 Duisburg, Germany}
\author{Birthe Papendell}
\affiliation{University of Duisburg-Essen, Faculty of Physics, Lotharstra\ss e 1, 47048 Duisburg, Germany}
\author{Klaus Hornberger}
\affiliation{University of Duisburg-Essen, Faculty of Physics, Lotharstra\ss e 1, 47048 Duisburg, Germany}

\begin{abstract}
Motivated by trapping and cooling experiments with non-spherical nanoparticles, we discuss how their combined rotational and translational quantum state is affected by the interaction with a gaseous environment. Based on the quantum master equation in terms of orientation-dependent scattering amplitudes, we evaluate the localization rate for gas collisions off an anisotropic van der Waals-type potential and for photon scattering off an anisotropic dielectric. We also show how pure angular momentum diffusion arises from these open quantum dynamics in the limit of weak anisotropies.
\end{abstract}

\maketitle

\section{Introduction}

Cavity experiments with levitated nanoparticles in high vacuum \cite{kiesel2013,asenbaum2013,Moore2014,millen2015,ranjit2015,Ranjit2016,jain2016,vovrosh2016} are promising candidates for ultra sensitive force sensors \cite{Marago2008,Olof2012,Ranjit2016}. They allow addressing fundamental questions of physics, for instance, the validity of the quantum superposition principle for massive particles \cite{chang2010,romeroisart2011,Arndt2014testing,Bateman2014,goldwater2015,li2016} or the form of the dispersion interaction \cite{emig2007}. For the interpretation of such coherence experiments it is essential to quantify and understand all relevant sources of environmental decoherence, most prominently collisions with a background gas \cite{hornberger2003a} or Rayleigh scattering of photons \cite{jain2016}.

Recently, there has been some effort to exploit that anisotropically shaped nanoparticles enhance the interaction with the cavity field \cite{chang2012,kuhn2015,hoang2016,stickler2016}. While such nanoparticles can be well controlled in solution and low vacuum \cite{Paterson2001,Bonin2002,Jones2009,Tong2010,Padgett2011,arita2013,Brzobohaty2015,Simpson2016,spesyvtseva2016,irrera2016}, coherence experiments involving the orientational degrees of freedom \cite{bhattacharya2007,romero2010,shi2013,shi2015,stickler2016,hoang2016} are still pending. Understanding the spatio-orientational decoherence processes in such experiments with ultra-cold anisotropic nanoparticles is a prerequisite for exploiting the quantum motion of the center of mass and the orientation.

In this article, we show how to account for the spatio-orientational decoherence of a nanoparticle interacting with a homogeneous background gas in a microscopically realistic fashion, and we specify the associated localization rate for the two most relevant decoherence scenarios: {\it (i)} gas collisions off an anisotropic van der Waals-type potential, and {\it (ii)} Rayleigh-Gans scattering off a nanorod  whose length is comparable to the laser wavelength. Both examples can be extended to other interaction potentials or particle shapes. Finally, we explain how the corresponding master equations give rise to pure angular momentum diffusion in the limit of weak anisotropies.

To account for the environment we use an extension of the quantum linear Boltzmann equation \cite{Hornberger2006,Hornberger2008}, which provides a microscopic Markovian description of the quantum dynamics of a particle propagating in a homogeneous gas in terms of scattering amplitudes. The resulting master equation yields a valid discription as long as the correlations between successive collisions are negligible so that a Markovian formulation in terms of a quantum dynamical semigroup is possible \cite{breuer}. Non-Markovian effects \cite{breuer2016} would become important if the gas cannot be considered ideal. Instead of fully accounting for the internal degrees of freedom \cite{smirne2010}, we exploit the simplifying fact that the orientation of a large nanoparticle is approximately constant during the scattering interaction with a single gas particle \cite{Bennewitz1964}. Thus, the orientation enters only in parametric fashion,  which allows us to give closed form expressions for the spatio-orientational localization rate. This separation of time scales proved valid already for atom-molecule scattering experiments \cite{Bennewitz1964}, and was used to describe the orientationally averaged center-of-mass decoherence of polar molecules \cite{walter2016} and the purely orientational decoherence \cite{timothesis} in an isotropic environment \cite{zhong2016}. We will also show how this treatment can be naturally extended to a gas of photons, as required to describe decoherence by thermal radiation.

\section{Ro-translational master equation in the monitoring approach}

It is our aim to formulate the Markovian quantum master equation for the ro-translational state operator $\rho$ of an anisotropic nanoparticle interacting with a homogeneous gas of structureless particles of density $n_{\rm g}$. We adapt the monitoring approach \cite{hornberger2007,Hornberger2008}, which is based on disentangling the state-dependent rate of collisions from the effect of a single scattering event. This brings into play the scattering operator $\sS$ of a single collision and the corresponding rate operator $\mathsf{\Gamma}$, both acting on the Hilbert space of the relative center of mass coordinate (operators are denoted by sans-serif characters). Using the exact scattering amplitudes one can thus account for the environmental collisions non-perturbatively within a Markovian framework.

Since the rotation period of a nanoparticle of mass $M$ is typically much longer than the interaction time with a single gas particle of mass $m \ll M$, the scattering operator $\sS$ and the rate operator $\mathsf{\Gamma}$ are diagonal in the orientational degrees of freedom $\Omega$ (sudden approximation) \cite{Bennewitz1964,Stickler2015b,walter2016,zhong2016}. Thus, the molecule's orientation enters the scattering amplitude $f({\bf p}_{\rm f},{\bf p}_{\rm i}; \Omega)$ (describing a collision with initial and final relative momentum ${\bf p}_{{\rm i}}$ and ${\bf p}_{{\rm f}}$, respectively) only in parametric fashion and the quantum master equation can be derived by repeating the steps demonstrated in \cite{Hornberger2008} but now including the orientational degrees of freedom.

We note that in the case of coherence experiments with rapidly rotating small molecules \cite{milner2014,zastrow2014}, the scattering operator and the rate operator are approximately diagonal in the angular momentum basis \cite{Stickler2015b}, giving rise to rotational decoherence \cite{adelswaerd2003,adelswaerd2004}.

\subsection{Monitoring Master Equation}

The starting point for deriving the master equation for the ro-translational state $\rho$ of the nanoparticle in the monitoring approach is the general expression
\be \label{eq:mon}
\partial_t \rho = -\frac{i}{\hbar} \left [ \sH, \rho \right ] + \cR \rho + \cL \rho,
\ee
where $\sH$ is the free Hamiltonian and the two superoperators in \eqref{eq:mon} are given as
\begin{subequations}
\be
\cR \rho = i \mathrm{tr}_{\rm g} \left ( \left [ \mathsf{\Gamma}^{1/2} \mathrm{Re}(\sT) \mathsf{\Gamma}^{1/2}, \rho \otimes \rho_{\rm g} \right ] \right ),
\ee
\bea
\cL \rho & = & \mathrm{tr}_{\rm g} \left ( \sT \mathsf{\Gamma}^{1/2}\rho \otimes \rho_{\rm g} \mathsf{\Gamma}^{1/2} \sT^\dagger \vphantom{\frac{1}{2}} \right.\notag \\
 && \left. - \frac{1}{2} \{ \rho \otimes \rho_{\rm g}, \mathsf{\Gamma}^{1/2} \sT^\dagger \sT \mathsf{\Gamma}^{1/2} \} \right ),
\eea
\end{subequations}
where $\sT$ is the non-trivial part of the scattering operator, $\sS = \mathsf{1} + i \sT$, and $\rho_{\rm g}$ is the state of the gas \cite{Hornberger2008}.

Hence, one considers the effect of a single collision on the initially uncorrelated total state operator $\rho \otimes \rho_{\rm g}$ properly accounting for the collision probability and subsequently traces out the gas. Since the gas is stationary and homogeneous, its state operator $\rho_{\rm g}$ is diagonal in the momentum basis; we denote the diagonal elements by $\mu({\bf p})$. As both the scattering operator $\sS$ and the rate operator $\mathsf{\Gamma}$ act only on the Hilbert space of relative coordinates, it is convenient to introduce the relative momentum vector
\be
\mathrm{rel} \left ( {\bf p}, {\bf P} \right ) = \frac{m_{\rm r}}{m} {\bf p} - \frac{m_{\rm r}}{M} {\bf P},
\ee
where $m_{\rm r} = m M / (m + M)$ is the reduced mass and ${\bf P}$ is the momentum of the nanoparticle.

The interaction with the environment enters the master equation in two ways: On the one hand, we obtain a momentum- and orientation-dependent energy shift $\sH_{\rm g}$ involving the real part of the forward scattering amplitude,
\bea \label{eq:eshift}
\sH_{\rm g} & = & - 2 \pi \hbar^2 \frac{n_{\rm g}}{m_{\rm r}} \int \dint{{}^3 {\bf p}} \mu({\bf p}) \notag \\
 && \times \mathrm{Re} \left [ f \left ( \mathrm{rel}({\bf p},\boldsymbol{\sP}),\mathrm{rel}({\bf p},\boldsymbol{\sP}); \pOmega \right )\right ],
\eea
to be added to the free ro-translational Hamiltonian $\sH$. On the other hand, the collisions with the gas particles give rise to a dissipator $\cL \rho$ of Lindblad form. Denoting by $\boldsymbol{\sR}$ the center-of-mass operator of the nanoparticle, the Lindlad operators can be given as
\bea \label{eq:lindblad}
\sL_{{\bf Q}{\bf p}} & = & \exp \left ( \frac{i}{\hbar} \boldsymbol{\sR} \cdot {\bf Q} \right ) \sqrt{\frac{n_{\rm g} m}{Q m_{\rm r}^2}} \notag \\
 && \times \mu^{1/2} \left [ {\bf p} + \left ( 1 + \frac{m}{M} \right ) \frac{{\bf Q}}{2} + \frac{m}{M} \boldsymbol{\sP}_\| \right ] \notag \\
 && \times f \left [ \mathrm{rel}({\bf p}, \boldsymbol{\sP}_\bot) - \frac{{\bf Q}}{2}, \mathrm{rel}({\bf p}, \boldsymbol{\sP}_\bot) + \frac{{\bf Q}}{2}; \pOmega \right ],
\eea
where $\boldsymbol{\sP}_\bot = \boldsymbol{\sP} - ({\bf Q} \cdot \boldsymbol{\sP})/Q^2$ is the component of the momentum operator orthogonal to ${\bf Q}$.

Thus, the quantum master equation~\eqref{eq:mon} for the ro-translational motion of the nanoparticle is
\bea \label{eq:boltzmann}
\partial_t \rho & = & -\frac{i}{\hbar} \left [ \sH + \sH_{\rm g} , \rho \right ] + \cL \rho,
\eea
with
\bea \label{eq:dissi}
 \cL \rho & = &  \int \dint{{}^3{\bf Q}} \int_{{\bf Q}^\bot} \dint{{}^2 {\bf p}}  \left ( \sL_{{\bf Q}{\bf p}} \rho \sL_{{\bf Q}{\bf p}}^\dagger \vphantom{\frac{1}{2}} \right. \notag \\
  && \left. - \frac{1}{2} \left \{ \rho, \sL_{{\bf Q}{\bf p}}^\dagger \sL_{{\bf Q}{\bf p}} \right \} \right ),
\eea
where ${\bf Q}^\bot$ denotes the plane orthogonal to ${\bf Q}$. This master equation describes the ro-translational dynamics of the nanoparticle in a low-pressure ideal gas, for which the correlations between successive collisions are negligible and the Markov approximation is justified. It follows from the form of the Lindblad operators \eqref{eq:lindblad}, that the dissipator in \eqref{eq:dissi} describes center-of-mass decoherence and dissipation \cite{Hornberger2008} as well as localization in the particle's orientation.

\subsection{Decoherence in the Configuration Coordinates}

Noting that the levitated nanoparticles used for metrology are typically much heavier than the gas atoms, $m / M \ll 1$, one can simplify the master equation~\eqref{eq:boltzmann}. The energy shift \eqref{eq:eshift} gets diagonal in the orientational degrees of freedom $\Omega$ and thus turns into a constant for isotropic gas distributions, while the Lindblad operators become diagonal in position ${\bf R}$ and orientation $\Omega$. The resulting superoperator describes decoherence in the configuration space and reads
\bea \label{eq:lcd}
\cL_{\rm c} \rho & = & \frac{n_{\rm g}}{m} \int \dint{{}^3{\bf p}} \int_{S_2} \dint{{}^2 {\bf n}'} p \mu( {\bf p}) \left [ \sA_{{\bf n}'{\bf p} } \rho \sA_{{\bf n}'{\bf p}}^\dagger \vphantom{\frac{1}{2}} \right. \notag \\
 && \left. - \frac{1}{2} \left \{ \rho, \sA_{{\bf n}'{\bf p} }^\dagger \sA_{{\bf n}'{\bf p} } \right \} \right ],
\eea
with the Lindblad operators
\be \label{eq:lindblad2}
\sA_{{\bf n}'{\bf p}} = e^{i \boldsymbol{\sR} \cdot({\bf p} - p {\bf n}')/\hbar} f(p {\bf n}', {\bf p}; \pOmega),
\ee
{\it i.e.}, ${\bf n}'$ denotes the direction into which the gas particle is scattered. If the center-of-mass degrees of freedom are traced out from Eq.~\eqref{eq:lcd}, one obtains the generator of pure orientational decoherence, which was recently discussed for isotropic gas distributions \cite{zhong2016}. Alternatively, tracing out the orientational degrees of freedom gives the orientation-averaged master equation for anisotropic point-like particles \cite{walter2016}. 

In the case that the gas distribution is isotropic, the configuration-space matrix elements of the superoperator \eqref{eq:lcd} can be written in the compact multiplicative form
\bea
\matel{{\bf R} \Omega}{\cL_{\rm c} \rho}{{\bf R}' \Omega'} & = &  - \left[ F_{\Omega \Omega'}({\bf R} - {\bf R}') - i G_{\Omega \Omega'}({\bf R} - {\bf R}') \right ] \notag \\
 && \times \matel{{\bf R} \Omega}{\rho}{{\bf R}' \Omega'},
\eea
by defining the localization rate
\bea \label{eq:locrate}
F_{\Omega \Omega'}({\bf R}) & = & \frac{n_{\rm g}}{2 m} \int_0^\infty \dint{p} p^3  \mu(p) \int_{S_2} \dint{{}^2{\bf n}} \dint{{}^2{\bf n}'} \notag \\
 && \times \left \vert f(p{\bf n}',p {\bf n}; \Omega) e^{i p {\bf R} \cdot ({\bf n} - {\bf n}')/\hbar} \right. \notag \\
  && \left. - f(p{\bf n}',p {\bf n}; \Omega') \vphantom{e^{i p {\bf R} \cdot ({\bf n} - {\bf n}')/\hbar}} \right \vert^2,
\eea
and the frequency
\bea \label{eq:rotrate}
G_{\Omega \Omega'}({\bf R}) & = & \frac{n_{\rm g}}{m} \int_0^\infty \dint{p} p^3  \mu(p) \int_{S_2} \dint{{}^2{\bf n}} \dint{{}^2{\bf n}'} \notag \\
 && \times \mathrm{Im} \left [ f(p{\bf n}',p {\bf n}; \Omega)f^*(p{\bf n}',p {\bf n}; \Omega') \vphantom{e^{i p {\bf R} \cdot ({\bf n} - {\bf n}')/\hbar}} \right. \notag \\
  && \left. \times e^{i p {\bf R} \cdot ({\bf n} - {\bf n}')/\hbar} \right ].
\eea
The localization rate is always positive and thus determines the time scale at which the spatio-orientational coherences decay. It vanishes for the diagonal elements ${\bf R} = {\bf R}'$ and $\Omega = \Omega'$. Note that even for isotropic gas distributions, the localization rate \eqref{eq:locrate} is in general not only a function of the distance $\vert {\bf R} - {\bf R}' \vert$, but of the distance vector ${\bf R} - {\bf R}'$ due to the anisotropic interaction potential.

In what follows, we will specify the localization rate \eqref{eq:locrate} for two common scenarios, scattering of gas particles off an anisotropic van der Waals-type interaction potential and light scattering off a dielectric nanorod.

\section{Anisotropic van der Waals Scattering}

We proceed to calculate explicitly the localization rate \eqref{eq:locrate} for a homogeneous potential with $\cos^2 \Theta$-anisotropy, where $\Theta$ is the angle between the principle axis of inertia of a symmetric top and the direction of incidence,
\be \label{eq:pot}
V(r, \cos \Theta) = - \frac{C}{r^s} ( 1 + a \cos^2 \Theta ),
\ee
Relevant examples for potentials of this type include the dipole-induced dipole interaction and the anisotropic van der Waals interaction \cite{StoneBook}. The dipole-induced dipole interaction, for instance, is specified by $s = 6$, $a = 3$, and $C = \alpha_0 d_0^2/32 \pi^2 \eps_0^2$, where $\alpha_0$ denotes the gas particle polarizability and $d_0 {\bf m}(\Omega)$ is the permanent electric dipole moment, {\it i.e.}, $\cos \Theta = {\bf m}(\Omega) \cdot {\bf r}/r$.

In order to evaluate the localization rate \eqref{eq:locrate}, the orientation-dependent scattering amplitude $f({\bf p},{\bf p}'; \Omega)$ is required. While the total scattering cross section can be evaluated by using the optical theorem together with Schiff's formula as obtained in the eikonal approximation \cite{Bennewitz1964},
\bea \label{eq:sigmatot}
\sigma_a(p{\bf n};\Omega) & = & 2 \pi \sin \left ( \frac{\pi}{2} \frac{s - 3}{s-1} \right ) \Gamma \left ( \frac{s -3}{s-1} \right ) \notag \\ 
 && \!\!\!\!\!\!\!\! \times \left ( \frac{\sqrt{\pi} m C}{\hbar p} \frac{\Gamma [ (s -1)/2]}{\Gamma(s/2)} \right )^{2 / (s-1)}\left ( 1 + \frac{a (s - 1)}{2 s} \right. \notag \\
  &&\!\!\!\!\!\!\!\!  \left. - \frac{a(s-3)}{2 s} ({\bf n} \cdot {\bf m}(\Omega))^2 \right )^{2/(s-1)},
\eea
a closed expression for the scattering amplitude can only be given for small-angle scattering \cite{walter2016},
\be \label{eq:scattamp}
f(p{\bf n'},p {\bf n};\Omega) \simeq f_{\rm fwd}(p{\bf n}; \Omega) e^{- \vert {\bf n} \times {\bf n}' \vert^2 \chi_a \left ( p {\bf n} ;\Omega \right ) }.
\ee
It includes the forward scattering amplitude
\be \label{eq:f0}
f_{\rm fwd}(p{\bf n};\Omega) = \frac{p \sigma_a(p {\bf n}; \Omega)}{4 \pi \hbar \cos [ \pi / (s - 1)]} \exp \left ( \frac{i \pi}{2} \frac{s - 3}{s-1} \right ),
\ee
and the function
\bea \label{eq:expfunc}
\chi_a(p {\bf n}; \Omega) & = & \left ( \frac{p}{2 \sqrt{2} \hbar} \right )^2 \frac{\sigma_a(p {\bf n}; \Omega )}{2 \pi \cos[\pi/(s-1)]} \Gamma \left (\frac{s-5}{s-1} \right ) \notag \\
 && \times \Gamma^{-2} \left (\frac{s-3}{s-1} \right ) \exp \left ( - \frac{i \pi}{s-1} \right ).
\eea
We emphasize that the small angle approximation \eqref{eq:scattamp} is usually sufficient to describe decoherence because hard collisions, $\vert {\bf n} \times {\bf n}' \vert \lesssim 1$, destroy the coherence completely, such that the details of the scattering amplitude do not matter. In contrast, the total scattering rate is evaluated by the optical theorem and thus without using the small angle approximation \cite{walter2016}.

\begin{figure}
\centering
\includegraphics[width = 80mm]{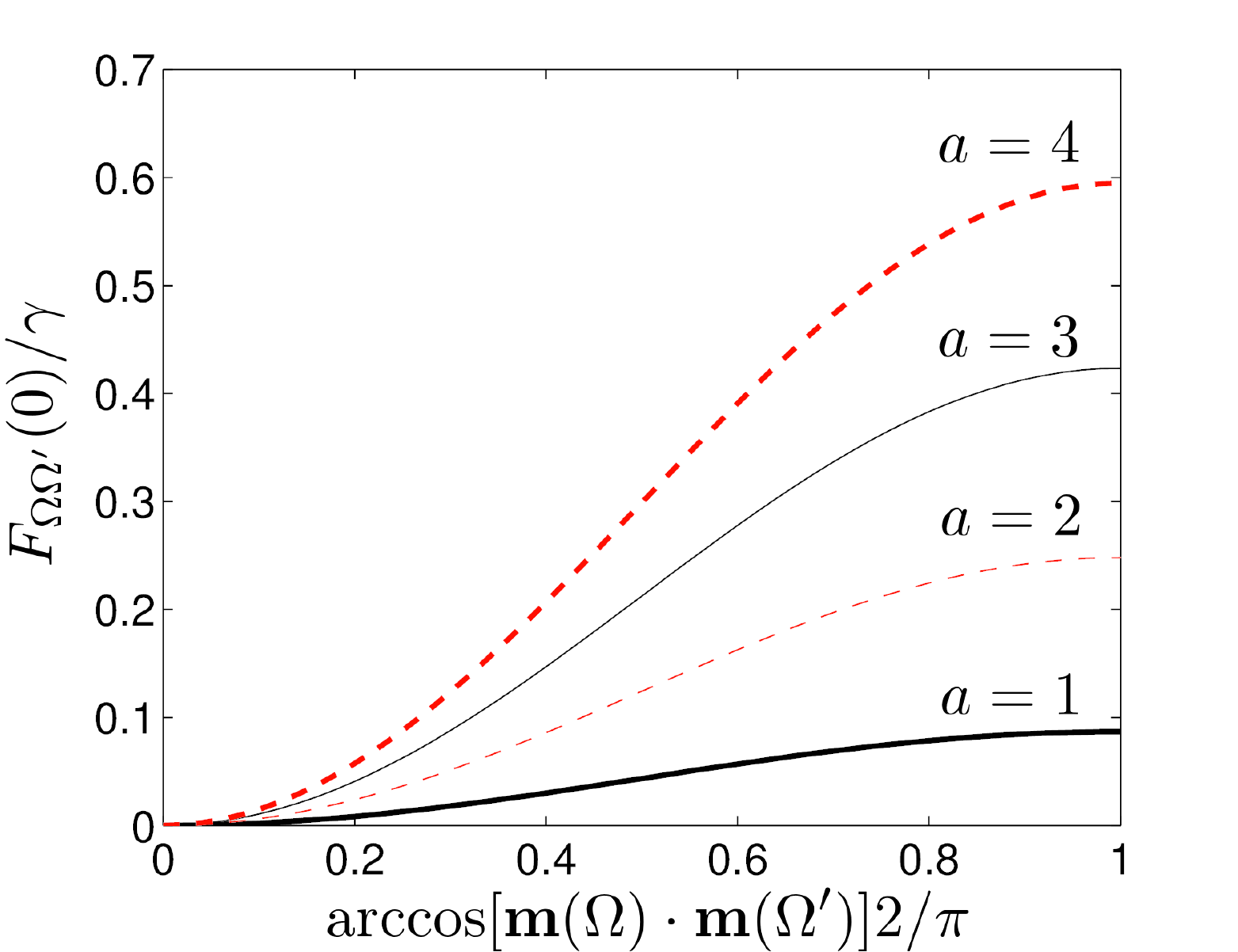}
\caption{Orientational localization rate $F_{\Omega \Omega'}(0)$ of a polar particle ($d_0 = 5$ D) interacting with a room-temperature gas of Helium atoms ($\alpha_0 / 4 \pi \eps_0 = 0.2$ \AA${}^3$). The localization rate is depicted in units of the rate \eqref{eq:gamma} and as a function of the angle between the orientations ${\bf m}(\Omega)$ and ${\bf m}(\Omega')$ for different anisotropies $a$.} \label{fig:locrate1}
\end{figure}

It follows from the scattering amplitude \eqref{eq:scattamp} that the orientational localization rate, Eq.~\eqref{eq:locrate} with ${\bf R} = 0$, depends only on the angle between the two different orientations of the dipole, {\it i.e.}, it depends only on ${\bf m}(\Omega) \cdot {\bf m}(\Omega')$. As expected, the orientational localization rate vanishes for identical orientations and is strictly increasing for increasing angle between the orientations. This can also be observed from Fig. \ref{fig:locrate1}, where we depict $F_{\Omega \Omega'}(0)$ as a function of the angle between ${\bf m}(\Omega)$ and ${\bf m}(\Omega')$ for different anisotropies $a$ of the interaction potential \eqref{eq:pot}. The localization rate increases with increasing anisotropy and it will be demonstrated below that for small anisotropies it is proportional to the squared sine of the angle between the two orientations. Finally, we remark that for combined spatio-orientational superpositions the localization rate \eqref{eq:locrate} depends not only on the distance $R$ and on the angle between ${\bf m}(\Omega)$ and ${\bf m}(\Omega')$, but also on the angles between ${\bf R} / R$ and ${\bf m}(\Omega)$ as well as ${\bf m}(\Omega')$.

\section{Photon Scattering}

As a second example, we consider Rayleigh-Gans scattering off a thin dielectric rod of length $\ell$, radius $a_0$ and permittivity $\eps_{\rm r}$, noting that the results presented here can be easily extended to other shapes, such as thin discs \cite{schiffer79,stickler2016}. We assume that the rod is illuminated by a linearly-polarized plane laser wave, ${\bf E}({\bf r}) = E_0 \boldsymbol{\epsilon}_{\rm p} e^{i k {\bf n} \cdot {\bf r}}$ (with ${\bf n} \cdot \boldsymbol{\epsilon}_{\rm p} = 0$), of wavelength $2 \pi / k$. Such a situation may arise in optomechanical experiments, where a laser field is used to manipulate or track the rod's dynamics \cite{kuhn2015}. For thin rods, $k^2 a_0^2 (\eps - 1) < 1$, the field strength is nearly constant along the width of the rod and the internal polarization field approximately assumes the position dependence of the external field (generalized Rayleigh-Gans approximation) \cite{schiffer79}. Knowledge of the internal polarization field then allows one to deduce the interaction potential as well as the master equation \cite{pflanzer2012,stickler2016}. The standard way of deriving the latter would consist of coupling the combined electric field-rod state to the infinite bath of vacuum modes and tracing them out in the Born-Markov approximation. In what follows, we will demonstrate that one can determine Lindblad operators of Rayleigh-Gans scattering straightforwardly from the quantum master equation~\eqref{eq:boltzmann} by using the scattered light field.

\subsection{Rayleigh-Gans Scattering Amplitude}

The Rayleigh-Gans scattering amplitude is a vectorial quantity since it depends on the initial and final polarization state of the light \cite{jackson1999}. Noting that the photon energy is independent of the polarization and that the state operator $\rho_{\rm g}$ is diagonal in the polarization, it is straightforward to adapt the master equation~\eqref{eq:boltzmann}. The Lindblad operators \eqref{eq:lindblad2} now depend explicitly on the polarization state of the scattered photon and there is an additional sum over the polarization in the superoperator \eqref{eq:dissi}. Moreover, the forward scattering amplitude entering the coherent energy shift \eqref{eq:eshift} is given by the vector scattering amplitude evaluated in the direction of the incoming polarization, which is familiar from optics \cite{jackson1999}. 

The electric field scattered from a thin rod at the origin oriented into direction ${\bf m}(\Omega)$ can be expressed in the far-field as \cite{schiffer79}
\be \label{eq:scattfield}
{\bf E}_{\rm sc}({\bf r}) = E_0 {\bf F}(p {\bf r}/r, p {\bf n} ;\Omega) \frac{e^{i k r}}{r},
\ee
with momentum $p = \hbar k$. Here, the vector scattering amplitude is given by
\bea \label{eq:ampray}
{\bf F}(p {\bf n}', p {\bf n}, \Omega) & = & - \frac{V_0 \chi_\| k^2}{4 \pi} {\bf n}' \times \left ( {\bf n}' \times {\bf u}(\Omega) \right ) \notag \\
 && \times {\rm sinc}\left [ \frac{k \ell}{2} {\bf m}(\Omega) \cdot ({\bf n} - {\bf n}') \right ],
\eea
where ${\rm sinc} (x) = \sin x / x$ and $V_0 = \pi \ell a_0^2$ is the volume of the rod. In addition, we defined the (unnormalized) direction of the internal polarization field,
\be
{\bf u}(\Omega) =  \frac{\chi_\bot}{\chi_\|} \boldsymbol{\epsilon}_{\rm p} + \frac{\Delta \chi}{\chi_\|} ({\bf m}(\Omega) \cdot \boldsymbol{\epsilon}_{\rm p}) {\bf m}(\Omega).
\ee
Here, $\boldsymbol{\epsilon}_{\rm p}$ is the polarization vector of the incoming laser light, $\chi_\|$ and $\chi_\bot$ denote the elements of the susceptibility tensor parallel and orthogonal to the rod's symmetry axis, respectively, and the susceptibility anisotropy is $\Delta \chi = \chi_\| - \chi_\bot$. In the case of a thin rod, the diagonal elements of the body-fixed susceptibility tensor can be given explicitly as $\chi_\| = \eps_{\rm r} - 1$ and $\chi_\bot = 2 (\eps_{\rm r} - 1) / (\eps_{\rm r} + 1)$ \cite{dehulst}.

The scalar scattering amplitude for scattering from polarization state $\boldsymbol{\epsilon}_{\rm p}$ into polarization state $\boldsymbol{\epsilon}_{\rm p}'$ is given by ${\bf F}(p {\bf n}, p {\bf n'}, \Omega) \cdot \boldsymbol{\epsilon}_{\rm p}'$. Hence, the forward scattering amplitude is the projection of ${\bf F}(p {\bf n}, p {\bf n'}, \Omega)$ onto $\boldsymbol{\epsilon}_{\rm p}$, and the modulus squared of the scattering amplitude \eqref{eq:ampray} gives the orientation-dependent differential scattering cross section \cite{jackson1999}. It is well known \cite{schiffer79,jackson1999} that the optical theorem does not apply to the scattering amplitude \eqref{eq:ampray} due to the Rayleigh-Gans approximation made in its derivation, and, thus, the total scattering cross section must be evaluated by integrating the differential cross section. However, this poses no problem if the approximations are justified. For instance, relation \eqref{eq:scattfield} predicts the scattering signal off thin silicon nanorods in a standing wave laser field with remarkable accuracy \cite{kuhn2015}.

\subsection{Photon Scattering Master Equation}

Inserting the scattering amplitude \eqref{eq:ampray} into the master equation of configurational decoherence \eqref{eq:lcd} and summing over the polarization directions $\boldsymbol{\epsilon}_{{\bf n}'s}$ of the scattered photon gives the required master equation of Rayleigh-Gans scattering. Since we consider a single running wave mode, the momentum distribution is $\mu({\bf p}) = \delta ({\bf p} - \hbar k {\bf n} )$. The density $n_{\rm g} = \vert b \vert^2 / V_{\rm m}$ is the total number of photons divided by the (large) mode volume  $V_{\rm m}$  of the laser beam and we replace $p / m$ by the speed of light $c$. Carrying out the integral over incoming momenta, one obtains
\bea \label{eq:lindblad3}
\cL_{\rm R} \rho & = & \gamma_0 \vert b \vert^2 \sum_{s = 1,2} \int_{S_2} \frac{\dint{{}^2{\bf n}'}}{4 \pi} \left [ \sB_{{\bf n}'s} \rho \sB_{{\bf n}'s}^\dagger \vphantom{\frac{1}{2}} \right. \notag \\
&& \left. - \frac{1}{2} \left \{ \rho, \sB_{{\bf n}'s}^\dagger \sB_{{\bf n}'s} \right \} \right ],
\eea
with the Rayleigh-Gans scattering rate $\gamma_0 = c V_0^2 \chi_\|^2 k^4/ 6 \pi V_{\rm m}$ and the Lindblad operators
\bea \label{eq:rayscatt}
\sB_{{\bf n}'s} & = & \sqrt{\frac{3}{2}} e^{i k ({\bf n} - {\bf n}') \cdot \boldsymbol{\sR}} {\boldsymbol{\epsilon}}_{{\bf n}'s} \cdot {\bf u}(\pOmega) \notag \\
 && \times \mathrm{sinc} \left [ \frac{k \ell}{2} {\bf m}(\pOmega) \cdot ({\bf n} - {\bf n}') \right ].
\eea
Equation~\eqref{eq:lindblad3} induces spatio-orientational localization of the rod's quantum state because the Lindblad operators \eqref{eq:rayscatt} are diagonal in position ${\bf R}$ and orientation $\Omega$, $\sB_{{\bf n}'s} = B_{{\bf n}'s}(\boldsymbol{\sR},\pOmega)$. We note that an analogous form of the Rayleigh-Gans scattering operator \eqref{eq:rayscatt} for thin rods and disks was derived for a standing-wave cavity mode by coupling the whole system to the bath of vacuum modes \cite{stickler2016} in a more tedious calculation.

The Lindblad operators for a polarizable point particle are recovered in the limit of vanishing rod length, $k \ell \to 0$, and in the limit of an isotropically polarizable nanoparticle, $\Delta \chi / \chi_\| \to 0$. Moreover, as will be discussed below, in the limit of small particles, $k \ell \ll 1$, the Lindblad operators \eqref{eq:rayscatt} describe angular momentum diffusion. 

The anisotropy of the photonic momentum distribution $\mu({\bf p})$ gives rise to an orientation-dependent energy shift \eqref{eq:eshift}  determined by the forward scattering cross section. Using the same replacements that lead to \eqref{eq:rayscatt}, Eq.~\eqref{eq:eshift} yields the laser potential with coupling frequency $U_0 = - \omega \chi_\| V_0 / 2 V_{\rm m}$,
\be
H_{\rm L}(\pOmega) = \hbar U_0 \vert b \vert^2 \left [ \frac{\chi_\bot}{\chi_\|} + \frac{\Delta \chi}{\chi_\|} ({\bf m}(\pOmega) \cdot \boldsymbol{\epsilon}_{\rm p})^2 \right ].
\ee
It serves to align the rod with the polarization axis $\boldsymbol{\epsilon}_{\rm p}$. This potential can be obtained alternatively by integrating the time-averaged potential energy density $-{\bf P} \cdot {\bf E}^* / 4$, where ${\bf P}$ denotes the polarization field within the rod \cite{pflanzer2012,stickler2016}.

The scattering rate and the coupling frequency $U_0$ can be expressed independently of the mode volume $V_{\rm m}$ by using that the classical field strength is related to the photon number by $E_0 = \sqrt{2 \hbar \omega / \eps_0 V_{\rm m}} b$. This leads to the replacements $\gamma_0 \vert b \vert^2 = \eps_0 E_0^2 \chi_\|^2 V_0^2 k^3/ 12 \pi \hbar$ and $\hbar U_0 \vert b \vert^2 = - \eps_0 \chi_\| V_0 E_0^2 / 4$.

\subsection{Photon Scattering Localization Rate}

The spatio-orientational localization rate \eqref{eq:locrate} of photon scattering follows from the master equation~\eqref{eq:lindblad3} as
\bea \label{eq:locrayleigh}
F_{\Omega \Omega'}({\bf R} - {\bf R}') & = & \frac{\gamma_0 \vert b \vert^2}{2} \sum_{s = 1,2} \int_{S_2} \frac{\dint{{}^2{\bf n}'}}{4 \pi} \left \vert B_{{\bf n}'s}({\bf R},\Omega) \right. \notag \\
 && \left. - B_{{\bf n}'s}({\bf R}',\Omega') \right \vert^2.
\eea
In contrast to the anisotropic van der Waals-type interaction discussed in the previous section, the orientational localization rate $F_{\Omega \Omega'}(0)$ depends on both orientations ${\bf m}(\Omega)$  and ${\bf m}(\Omega')$ individually since the field polarization $\boldsymbol{\epsilon}_{\rm p}$ and the propagation direction ${\bf n}$ of the laser define distinguished directions.

However, if the nanoparticle is illuminated incoherently by plane waves from a random direction with random polarization, as is the case in black body radiation, Eq.~\eqref{eq:locrayleigh} must be averaged over the direction of the incoming light beam ${\bf n}'$, over the polarization direction $\boldsymbol{\epsilon}_{\rm p}$, and over the momentum distribution $\mu(\hbar k)$. The coherent energy shift can then be ignored and carrying out the angular integration shows that the resulting orientational localization rate depends only on the angle between ${\bf m}(\Omega)$ and ${\bf m}(\Omega')$. Such a situation might arise if stray light cannot be avoided experimentally or if thermal radiation plays a role. The explicit form of the localization rate in the case of black body illumination is given below for small anisotropies.

\begin{figure}
\centering
\includegraphics[width = 80mm]{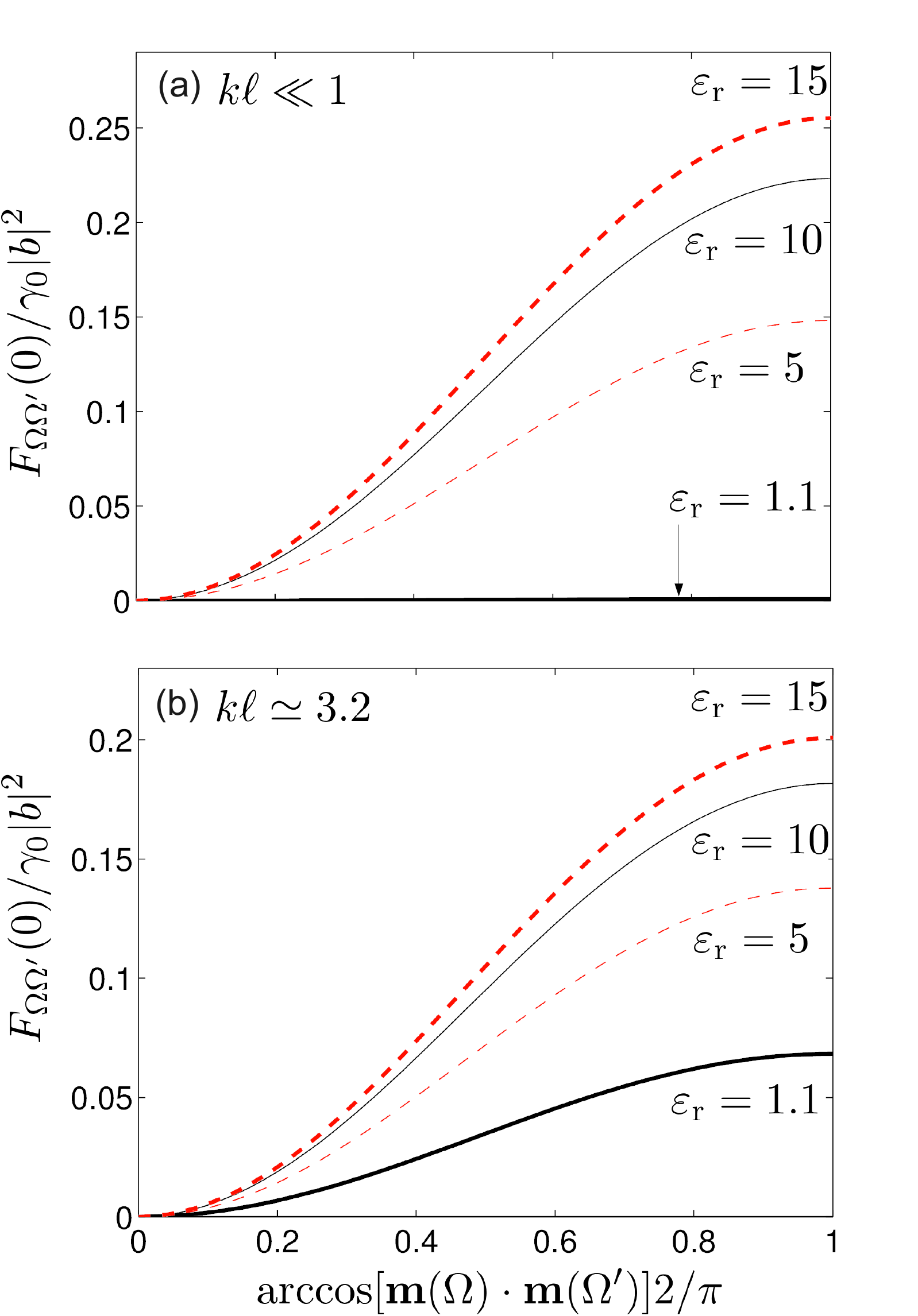}
 \caption{Orientational localization rate $F_{\Omega \Omega'}(0)$ of Rayleigh-Gans scattering from a monochromatic, isotropic and non-polarized radiation source. The localization rate is depicted in units of the Rayleigh-Gans scattering rate $\gamma_0$, as a function of the angle between the orientations ${\bf m}(\Omega)$ and ${\bf m}(\Omega')$, and for four different relative dielectric permittivities. In panel (a), the laser wavelength is assumed to be much larger than the rod length, $k \ell \ll 1$, such that the orientational decoherence depends only on the susceptibility tensor of the rod, Eq.~\eqref{eq:rayscatt}. In panel (b), the laser wavelength and the rod length are of comparable magnitude, $\lambda = 1.56$ $\mu$m and $\ell = 0.8$ $\mu$m.} \label{fig:locrate2}
\end{figure}

In Fig. \ref{fig:locrate2} we show the resulting localization rate in units of the Rayleigh-Gans scattering rate $\gamma_0$ for different dielectric permittivities $\eps_{\rm r}$, {\it i.e.}, different relative anisotropies $\Delta \chi / \chi_{\|}$. The radiation is assumed to be monochromatic but its polarization is uniformly distributed. Again, the orientational localization rate increases with increasing anisotropy and it vanishes in the limit that the particle is transparent, $\eps_{\rm r} = 1$, because then $\gamma_0 = 0$. We also show the influence of the particle extension, $k \ell \gtrsim 1$, on the orientational localization rate.

\section{Angular Momentum Diffusion}

We now proceed to discuss the master equations of gas and Rayleigh-Gans scattering \eqref{eq:lcd} and \eqref{eq:lindblad3} in the case of small anisotropies and demonstrate that in this limit both turn into the universal master equation of pure angular momentum diffusion for symmetric top nanoparticles. Focussing on the rotation state $\rho_{\rm r} = \mathrm{tr}_{\rm cm}(\rho)$ the master equation assumes the form
\bea \label{eq:diff}
\cL_{\rm d} \rho_{\rm r} & = & \frac{15 D}{2 \hbar^2} \int_{S_2} \frac{\dint{{}^2 {\bf n}}}{4\pi} \left ( \left [{\bf n} \cdot {\bf m}(\pOmega) \right ]^2 \rho_{\rm r} \left [ {\bf n} \cdot {\bf m}(\pOmega) \right ]^2 \right. \notag \\
 && \left. - \frac{1}{2} \left \{ \rho_{\rm r}, \left [{\bf n} \cdot {\bf m}(\pOmega) \right ]^4 \right \} \right ).
\eea
It depends on the details of the environmental interaction only through the positive angular momentum diffusion coefficient $D$, which will be evaluated below.

In the orientation basis we thus have $\matel{\Omega}{\cL_{\rm d} \rho_{\rm r}}{\Omega'} = - F_{\Omega \Omega'} \matel{\Omega}{\rho_{\rm r}}{\Omega'}$ with a localization rate
\be \label{eq:rateor}
F_{\Omega \Omega'} = \frac{D}{\hbar^2} \left \vert {\bf m}(\Omega) \times {\bf m}(\Omega') \right \vert^2,
\ee
that is manifestly positive and increases with the squared sine of the angle between the two orientations ${\bf m}(\Omega)$ and ${\bf m}(\Omega')$. This dependence on the squared sine is the orientational analogue to the squared distance scaling of the center of mass localization rate of momentum diffusion \cite{joos1985,joos2003,breuer} and was already found for thermal photon scattering off anisotropic point-like particles and for gas scattering off a Gaussian potential in the Born approximation \cite{zhong2016}.

\subsection{Time Evolution of Expectation Values}

In order to demonstrate that Eq.~\eqref{eq:diff} indeed generates angular momentum diffusion, we use Euler angles $\Omega = (\alpha,\beta,\gamma)$ in the $z$-$y'$-$z''$ convention \cite{edmonds1996,brink2002}. The canonically conjugate momentum operators are denoted by $\pop_\alpha$, $\pop_\beta$, and $\pop_\gamma$, respectively, and they obey the canonical commutation relations
\be
[{\bf m}(\pOmega), \pop_\alpha] = i\hbar \sin \upbeta {\bf e}_{\alpha}(\pOmega), \quad [{\bf m}(\pOmega),  \pop_\beta ] = i\hbar {\bf e}_\beta(\pOmega),
\ee
together with $[{\bf m}(\pOmega),\pop_\gamma] = 0$, because ${\bf m}(\Omega) = (\cos \alpha \sin \beta, \sin \alpha \sin \beta, \cos \beta)^T$ depends only on the azimuthal angle $\alpha$ and the polar angle $\beta$.

The space-fixed angular momentum operators of the rigid rotor are related to the Euler momentum operators by \cite{edmonds1996}
\begin{subequations} \label{eq:angmom}
\bea
\sJ_1 & = & - \left ( \frac{\cot \upbeta}{2}  \left \{\pop_\alpha,\cos \upalpha \right \} + \sin \upalpha \pop_\beta -\frac{\cos \upalpha}{\sin \upbeta} \pop_\gamma \right ) \\
\sJ_2 & = & - \left ( \frac{\cot \upbeta }{2} \left \{\pop_\alpha,\sin \upalpha \right \} - \cos \upalpha \pop_\beta -\frac{\sin \upalpha}{\sin \upbeta} \pop_\gamma \right ) \\
\sJ_3 & = & \pop_\alpha,
\eea
\end{subequations}
and they obey the commutation relations $[\sJ_i,\sJ_j] = i \hbar \eps_{ijk} \sJ_k$.

The time evolution of the expectation values of the angular momentum $\boldsymbol{\sJ} = (\sJ_1,\sJ_2,\sJ_3)^T$ is unaffected by the generator \eqref{eq:diff},
\be
\mathrm{tr} \left ( \boldsymbol{\sJ} \cL_{\rm d} \rho_{\rm r} \right ) = 0.
\ee
because $\mathrm{tr}(\pop_\alpha \cL_{\rm d} \rho_{\rm r}) = \mathrm{tr}(\pop_\beta \cL_{\rm d} \rho_{\rm r}) = \mathrm{tr}(\pop_\gamma \cL_{\rm d} \rho_{\rm r}) = 0$.

Thus, the angular momentum operator expectation value $\mitl{\boldsymbol{\sJ}}$ evolves as if orientational decoherence were not present. In a similar fashion, the effect of Eq.~\eqref{eq:diff} on the second moment of the angular momentum operator $\mitl{\boldsymbol{\sJ}^2}$ is
\be
\mathrm{tr} \left ( \boldsymbol{\sJ}^2 \cL_{\rm d} \rho_{\rm r} \right ) =4 D.
\ee
In particular, if the coherent time evolution is determined by the free rotational Hamiltonian $\sH_{\rm r}$, the first moment of the angular momentum operator is conserved, $[\sH_{\rm r}, \sJ_i] = 0$, while its second moment increases linearly in time
\be \label{eq:totangmom}
\mitl{\boldsymbol{\sJ}}_t = \mitl{\boldsymbol{\sJ}}_0,\quad \mitl{ \boldsymbol{\sJ}^2}_t = \mitl{\boldsymbol{\sJ}^2}_0 +4 D t.
\ee
This demonstrates that the generator \eqref{eq:diff} indeed describes angular momentum diffusion.

\subsection{Angular Momentum Distribution}

In order to calculate the time-dependent probability distribution of angular momenta, $\matel{jm}{\rho_{\rm r}(t)}{jm}$, we study the time evolution of the $j = 0$ angular momentum eigenstate, $\rho_{\rm r}(0) = \ketbra{j = 0,m= 0}{j = 0, m = 0}$. Since we are only interested in the effects of the diffusion master equation~\eqref{eq:diff} we neglect the free time evolution. Thus, one expects that the angular momentum populations are given in the semiclassical limit by a Gaussian probability density whose variance increases linearly with time. The initial orientational coherences of $\rho_{\rm r}(0)$ decay exponentially with the rate \eqref{eq:rateor} and the marginal probabilities $p_t(j) = \sum_m \matel{j m}{\rho_{\rm r}(t)}{j m}$ take on the form
\be \label{eq:pl}
p_t(j) = \frac{2 j + 1}{2} \int_0^\pi \dint{\theta} \sin \theta P_j(\cos \theta ) \exp \left ( - \frac{D t}{\hbar^2} \sin^2 \theta \right ).
\ee
A straightforward calculation exploiting the completeness of Legendre polynomials $P_j(\cos \theta)$ \cite{nist} allows one to demonstrate that the variance indeed increases linearly in time,
\be
\mitl{\boldsymbol{\sJ}^2} = \sum_{j = 0}^\infty j (j + 1) p_t(j) = 4 D t.
\ee

Moreover, in the semiclassical regime, $D t/\hbar^2 \gg 1$ and $j \gg 1$, only small angles $\theta \ll 1$ contribute to the integral \eqref{eq:pl} and one can extend the $\theta$-integration to infinity while replacing $\sin \theta \simeq \theta$. Further noting that $P_j(\cos \theta) \simeq J_0[ (j + 1/2) \theta]$ for $j \gg 1$ \cite{nist} shows that Eq.~\eqref{eq:pl} is, as expected, asymptotically equivalent to the Gaussian distribution of angular momentum eigenvalues,
\be
p_t(j) \simeq (2 j + 1) \frac{\hbar^2}{4 Dt} \exp \left [ - \frac{\hbar^2}{4 D t} \left ( j + \frac{1}{2} \right )^2 \right ].
\ee

\subsection{Classical Diffusion Equation}

For comparison, the classical diffusion equation for the linear rotor can be derived by expressing the dissipator~\eqref{eq:diff} in the quantum phase space of the rotation state \cite{fischer2013,timothesis} and then drawing the classical limit \cite{timothesis}. Denoting by $H_{\rm rot}(\Omega,p_\Omega)$ the Hamilton function, the evolution equation for the phase space distribution $f(\Omega,p_\Omega,t)$ reads
\be \label{eq:cldiff}
\partial_t f + \{ f, H_{\rm rot} \}_{\rm P} = D  \left ( \sin^2 \beta \partial_{p_\alpha}^2  +  \partial_{p_\beta}^2 \right ) f,
\ee
where $\{f,g \}_{\rm P} = \sum_i (\partial_{x_i}f \partial_{p_i} g - \partial_{x_i}g \partial_{p_i} f)$ denotes Poisson's bracket.

Rotational friction can be taken into account by adding the term $-(D/I k_{\rm B} T) ( \partial_{p_\alpha} p_\alpha + \partial_{p_\beta} p_\beta) f$. One can verify by direct calculation that the steady-state solution is then given by the Boltzmann distribution with energy $H_{\rm rot}$ and temperature $T$.

\subsection{Diffusion Coefficients}

We now determine the orientational diffusion coefficients of photon- and van der Waals-type scattering. Starting with the latter, we trace out the center-of-mass degrees of freedom in \eqref{eq:lcd} and expand the forward scattering amplitude \eqref{eq:scattamp} to lowest order in $a$. Comparing the resulting master equation with the diffusion master equation~\eqref{eq:diff} gives
\be
D = \frac{2 \gamma (\hbar a)^2}{15}.
\ee
with the rate
\bea \label{eq:gamma}
\gamma & = & \frac{n_{\rm g}}{2 m \hbar^2 \cos^2[\pi/(s-1)]} \left ( \frac{s-3}{s(s-1)} \right )^2 \notag \\
 && \times\int_0^\infty \dint{p} \int_{-1}^1 \dint{\xi}\mu(p) p^5 \sigma_0^2(p) \exp \left [ -2 (1 - \xi^2) \chi_0(p) \right ] \notag \\
  && \times \left \vert (1 - \xi^2) \chi_0(p) - 1 \right \vert^2.
\eea
If the gas is in thermal equilibrium, $\mu(p)$ is given by the Maxwell-Boltzmann distribution and the rate \eqref{eq:gamma} is a function of the gas temperature $T$ and the gas density $n_{\rm g}$.

In a similar fashion, the diffusion master equation due to non-polarized monochromatic Rayleigh-Gans scattering can be obtained by expanding the scattering amplitude \eqref{eq:ampray} to first order in the relative susceptibility anisotropy $\Delta \chi / \chi_\|$ and in the wave number $k \ell$. Thus, one obtains the diffusion coefficient
\be \label{eq:diffr1}
D_{\rm R} =   \gamma_0 \vert b \vert^2 \hbar^2 \left ( \frac{1}{3} \left ( \frac{\Delta \chi}{\chi_\|} \right )^2 + \frac{(k \ell)^4}{540} \right ).
\ee

If the particle is illuminated by a black body in thermal equilibrium at temperature $T$, the distribution of wavenumbers $k$ follows from Planck's law, $\mu(k) = k^2/ n_{\rm g} \pi^2 [ \exp ( \hbar c k / k_{\rm B} T) -1 ]$ with temperature $T$. The resulting diffusion coefficient can be obtained by using $\vert b \vert^2 / V_{\rm m} = n_{\rm g}$  in \eqref{eq:diffr1} and averaging over the momentum distribution $\mu(k)$,
\bea
D_{\rm bb} & = & 40 \frac{c (\hbar \chi_\| V_0)^2}{\pi^3} \left ( \frac{k_{\rm B} T}{\hbar c} \right )^7 \left [ \zeta(7) \left ( \frac{\Delta \chi}{\chi_\|} \right )^2 \right. \notag \\
 && \left. + 28 \zeta(11) \left ( \frac{k_{\rm B} T \ell}{\hbar c} \right )^4 \right ],
\eea
where $\zeta(\cdot)$ denotes the Riemann $\zeta$-function. Inserting the diffusion coefficient into \eqref{eq:rateor} gives the orientational localization rate of black body radiation. In particular, for $k_{\rm B} T \ell/\hbar c \to 0$ one obtains the orientational localization rate of point-like particles \cite{zhong2016}.

The remarkable fact that all master equations considered here show angular momentum diffusion for small anisotropies suggests that Eq.~\eqref{eq:diff} is of a universal form. We remark that a unitary version of such a universal master equation was discussed in \cite{timothesis}.

\vspace{0.3cm}

\section{Conclusions}

We presented and discussed the master equation of spatio-orientational decoherence of a nanoparticle interacting with a homogeneous background gas and provided microscopically realistic descriptions for two important scenarios: {\it (i)} the dynamics due to scattering off a generic anisotropic van der Waals-type potential and {\it (ii)} Rayleigh-Gans scattering off a nanorod. We hope that the framework and equations presented in this article  will contribute to understanding coherence experiments with optically levitated particles in high vacuum \cite{bhattacharya2007,shi2015,hoang2016,stickler2016}. It may also become relevant for  matter wave interferometry with orientational degrees of freedom \cite{shore2015,Stickler2015b} or dedicated collision experiments with atomic beams. Furthermore, the fact that angular momentum diffusion arises in the limit of small anisotropies in all considered scenarios, suggests that the derived angular momentum diffusion master equation is a universal form for symmetric-top particles. In this regime, the orientational coherences decay in proportion to the squared sine between the two orientations.


\end{document}